\documentclass{acm_proc_article-sp}
\usepackage[commentsnumbered,vlined,linesnumbered]{algorithm2e}
\usepackage[]{fontenc}
\usepackage{hyperref}
\usepackage{amsmath}
\usepackage{mathtools}
\usepackage{graphicx}
\graphicspath{{poster_graphics/}{graphics/}{paper_graphics/}{New_plots/}}
\usepackage{lipsum}
\usepackage{epsfig}
\usepackage{graphicx,epstopdf}
\DeclareGraphicsExtensions{.ps}
\usepackage{verbatim}
\usepackage{multicol}
\usepackage{subfigure}
\newcommand{\quotes}[1]{``#1''}

\makeatletter
\def\@copyrightspace{\relax}
\makeatother

\begin{document}

\title{On Low Overlap Among Search Results of Academic Search Engines }
\numberofauthors{1}
\author{
%
%
\alignauthor
Anasua Mitra, Amit Awekar\\
       \affaddr{Indian Institute of Technology, Guwahati, Assam, India}\\
       \email{anasua.mitra, awekar@iitg.ernet.in}
}
       
\maketitle

\begin{abstract}
Number of published scholarly articles is growing exponentially. To tackle this information overload, researchers are increasingly depending on niche academic search engines. Recent works have shown that two major general web search engines: Google and Bing, have high level of agreement in their top search results. In contrast, we show that various academic search engines have low degree of agreement among themselves. We performed experiments using 2500 queries over four academic search engines. We observe that overlap in search result sets of any pair of academic search engines is significantly low and in most of the cases the search result sets are mutually exclusive. We also discuss implications of this low overlap.
     
\end{abstract}




\section{Introduction}\label{Intro}
Number of published research papers approximately doubles after nine years\cite{ASI:ASI23329}. This robust trend which is consistently observed after the second world war can be attributed to many factors. Sheer number of researchers is increasing with improvements in academic and research infrastructure in countries such as China and India. Researchers increasingly face publish-or-perish pardigm in research universities and laboratories. Web based systems for paper submission, reviewing and publication have significantly brought down the timespan and cost of publishing a paper. As a result, large number of low quality and even predatory research conferences and journals have emerged.

It is impossible to manually keep track of all relevant literature for any research topic. For example, in 2016 more than 2000 papers were published with words \quotes{deep learning} mentioned in the title\footnote{Scource: Google Scholar}. Actual number of papers published in 2016 on deep learning will be far more than that. In response to this information overload, many academic search engines (ASEs) have been developed to find appropriate research articles. These ASEs differ in multiple aspects: broad vs. specific topic coverage, web crawl vs. curated databases as source of data, commercial vs. non profit. ASEs play significant role in deciding which research papers are read by researchers. Therefore it is necessary to study ASEs separately from general search engines.

\section{Related Work}\label{relWork}
Overlap in the coverage and search results of web search engines is well studied. During mid 90s, multiple search engines such as AltaVista, Excite, and Lycos were competing with each other. These search engines covered only about 3 to 4\% of the web\cite{lawrence1998searching} and overlap in their search results was as low as 1.4\%\cite{bharat1998technique}. By the year 2005, major search engines were indexing more than 60\% of the web, but overlap in their results was still only about 11\%. Currently, there are only two major web search engines: Google and Bing. Recent work by Agrwal et. al. has shown that both these web search engines have high level of agreement between them\cite{agrawal2015study}. This convergence can be attributed to multiple factors such as high coverage of both search engines, maturity in ranking algorithms, and possibility of copying results from each other. Bibliographic datasets have been shown to have low overlap in their coverage\cite{hood2003overlap}. Various works have tried to predict coverage of ASEs\cite{orduna2015methods}. However to the best of our knowledge, there is no existing work that systematically studies overlap in the search results of ASEs.

\section{Our Work}\label{ourWork}
Motivation for systematically studying agreement among multiple ASEs primarily came from our own experience of using ASEs for research literature review. Almost mutually exclusive results obtained from multiple ASEs, prompted us to verify whether it was just a coincidence. This observation was in sharp contrast with recent work by Agrawal et. al.\cite{agrawal2015study} where they observed strong agreement among web search engines.

  \begin{figure*}[t]
  \centering
    \includegraphics[width=1.0\textwidth, height=5cm]{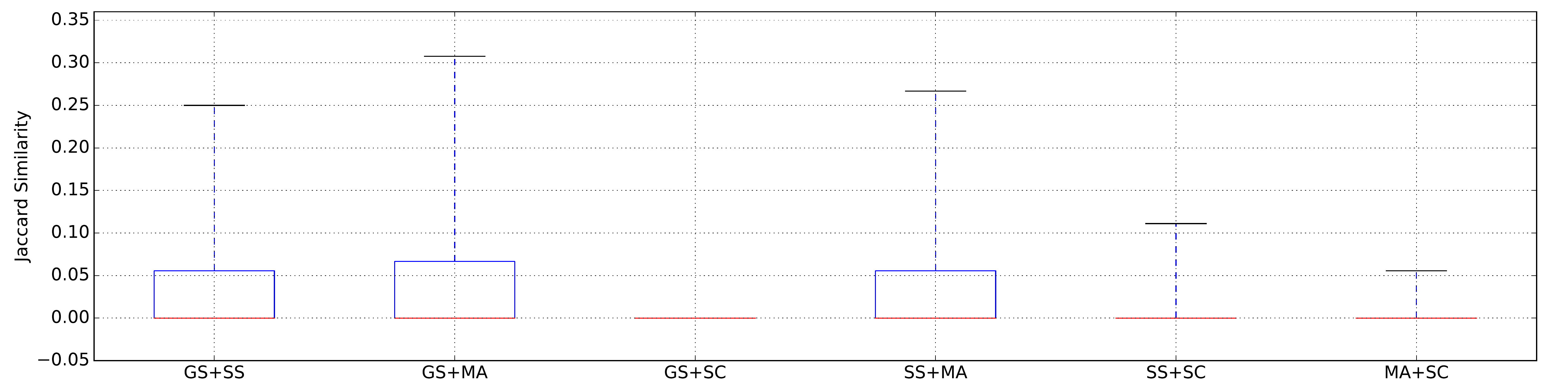}
    \caption{ Overlap in Search Result Sets of Academic Search Engines} 
    \label{KNN_list}
  \end{figure*}
  
We queried four ASEs: Google Scholar (GS)\footnote{\url{http://scholar.google.com/}}, Semantic Scholar (SS)\footnote{\url{http://www.semanticscholar.org/}}, Microsoft Academic (MA)\footnote{\url{https://academic.microsoft.com/}}, and Scopus (SC)\footnote{\url{https://www.scopus.com/home.uri}}. Main reason for choosing these ASEs was that they are popular in computer science and engineering domain. We collected over 2300 query terms from 2012 ACM Computing Classification System\footnote{\url{https://www.acm.org/publications/class-2012}}. This system arranges various computer science topics into a poly-hierarchy ontology. It sufficiently covers broad spectrum of topics in computer science from coarse to fine granularity. We also collected over 200 keywords from papers published in ACM SIGKDD 2016 conference.

We sent these  2500 queries to all four selected ASEs. For each ASE, we considered results only from the first page as users seldom go beyond the first page of search results. Various ASEs return different number of results on their first page. We looked at top eight results as each ASE returns at least eight results on the first page. We ignored the order in results and treated them as sets. If we consider order of results, then it will further drive down the similarity scores. We computed similarity between any two sets using the Jaccard similarity, that is ratio of size of intersection and union.

Please refer to Figure 1. X axis represents pairs of various ASEs. There are total six pairs as we considered four ASEs. Y axis represents Jaccard similarity for each pair using boxplots. Using 2500 queries, we obtained 2500 similarity scores for each ASE pair. Out of these scores, the figure depicts maximum (top black line), minimum (bottom black line), median (red line), 25 percentile, and 75 percentile (blue box) scores for each pair.

For all pairs, minimum score is always zero. It means that we have at least one query per pair such that their result sets are mutually exclusive. For all pairs, median score is also zero indicating that for most of the queries search result sets of ASEs are mutually exclusive. For all 2500 queries, intersection set of all four ASEs considered together was always empty. In other words, for each query, no research article appears in the top results list of all four ASEs. This shows strong diagreement among ASEs. Our queries covered topics of coarse as well as fine granularity. However, we did not see any correlation between topic granularity and overlap in search results.

GS and MA have comparable coverage of research literature (160 million and 150 million documents respectively). These two ASEs cover multiple topics including computer science. While SS covers only about 10 million documents, mostly from computer science. Compared to these three ASEs, SC has far small share of computer science documents. All our queries are from computer science domain.Therefore blue box comprising of 25 percentile and 75 percentile similarity scores is wide for pairs consisting of GS, MA, and SS. Where as pairs involving SC, the blue box almost flattens to line overlapping with minimum value of zero.

\section{Conclusion and Future Work} \label{conclude}
Unlike web search engines, ASEs do not have one or two dominant players. This has led to multiple systems that differ in coverage of research literature and ranking algorithms. Therefore overlap among search results of ASEs is signficantly low. As a result, users of ASEs have to look across multiple ASEs to find relevant research literature. We are working on extending this study in three ways. First, we are including more ASEs in the comparison. Second, we are using more diverse queries related to other subjects apart from just computer science. Third, we want to compare ASEs based on quality of search results.


\bibliographystyle{acm}
\bibliography{sample}  

\begin{thebibliography}{1}

\bibitem{agrawal2015study}
{\sc Agrawal, R., Golshan, B., and Papalexakis, E.}
\newblock A study of distinctiveness in web results of two search engines.
\newblock In {\em International Conference on World Wide Web\/} (2015),
  pp.~267--273.

\bibitem{bharat1998technique}
{\sc Bharat, K., and Broder, A.}
\newblock A technique for measuring the relative size and overlap of public web
  search engines.
\newblock {\em Computer Networks and ISDN Systems 30}, 1 (1998), 379--388.

\bibitem{ASI:ASI23329}
{\sc Bornmann, L., and Mutz, R.}
\newblock Growth rates of modern science: A bibliometric analysis based on the
  number of publications and cited references.
\newblock {\em Journal of the Association for Information Science and
  Technology 66}, 11 (2015), 2215--2222.

\bibitem{hood2003overlap}
{\sc Hood, W.~W., and Wilson, C.~S.}
\newblock Overlap in bibliographic databases.
\newblock {\em Journal of the American Society for Information Science and
  Technology 54}, 12 (2003), 1091--1103.

\bibitem{lawrence1998searching}
{\sc Lawrence, S., and Giles, C.~L.}
\newblock Searching the world wide web.
\newblock {\em Science 280}, 5360 (1998), 98--100.

\bibitem{orduna2015methods}
{\sc Orduna-Malea, E., Ayll{\'o}n, J.~M., Mart{\'\i}n-Mart{\'\i}n, A., and
  L{\'o}pez-C{\'o}zar, E.~D.}
\newblock Methods for estimating the size of google scholar.
\newblock {\em Scientometrics 104}, 3 (2015), 931--949.

\end{thebibliography}

\end{document}